\documentclass[prd,showpacs,showkeys]{revtex4}
\usepackage{graphicx,amsfonts,amssymb}

\begin{document}

\title{Finite Temperature Induced Fermion Number for Quarks in a Chiral Field}

\author{Gerald V. Dunne and Kumar Rao}

\affiliation{Department of Physics, University of Connecticut,  Storrs CT 06269-3046, USA\\}

\begin{abstract}

We compute the finite temperature correction to the induced fermion number for fermions coupled to a static SU(2) chiral background, using the derivative expansion technique. At zero temperature the induced fermion number is topological, being the winding number of the chiral background. At finite temperature however, higher order terms in the derivative expansion give nontopological corrections to the winding number. We use this result to show that the standard cancellation of the fractional parts of the fermion number inside and outside the bag in an SU(2) x SU(2) hybrid chiral bag model of the nucleon does not occur at nonzero temperature.

\end{abstract}

\maketitle

\section{Introduction}
When second quantized fermions interact with  a classical background field with a nontrivial topology (e.g, solitons, vortices, monopoles, Skyrmions etc) the resulting quantum states of the system can possess fractional eigenvalues. This intriguing phenomenon, known as fermion number fractionization, has applications ranging from particle, nuclear and atomic physics to condensed matter systems \cite{jr,gold,jaffe,jackiw,wilczek,niemi,heeger,amo}. At zero temperature, the fermion number of the vacuum is a topological quantity (up to spectral flow effects), and is related to the spectral asymmetry of the relevant Dirac operator, which counts the difference between the number of positive and negative energy states of the fermion spectrum. Rigorous mathematical results, such as index theorems and Levinson's theorem, show that the fractional part of the vacuum fermion number is a topological invariant; i.e, it depends only on the asymptotic values of the background field, and is invariant under local deformations of the background \cite{niemi,macwilczek,boyanovsky,farhi}. This topological character of the fermion number is important for various applications in model field theories, such as soliton models for the nucleon since it allows the fermion number to be kept fixed in a variational calculation that minimizes the energy \cite{ripka,bhaduri,weigel}. However, at finite temperature, the induced fermion number is  generically nontopological, and moreover is not a sharp observable \cite{dunneian,dunne,fluc,nonlinearsigma}. Several explicit examples of finite temperature fermion number in kink \cite{semenoff,dunneian}, sigma model \cite{dunneian,fluc} and monopole \cite{cp,dunneian,feinberg} backgrounds have been analyzed in detail. In this paper, we compute the finite temperature nontopological corrections to the induced fermion number for a (3+1) dimensional SU(2) chiral background, and show explicitly that the finite temperature correction is nontopological.

The model of second quantized fermions chirally coupled to classical, heavy scalar and pseudoscalar fields has applications in particle and nuclear physics, where it is used in models to describe the low energy spectrum of QCD such as in chiral quark soliton models of hadrons \cite{diakonov,ball,ripka} and in the hybrid chiral bag model of the nucleon \cite{jaffe,bhaduri}. At zero temperature, the induced fermion number of fermions interacting with a static chiral field coincides with the winding number of the field, and this result may be interpreted in terms of a topological current density. Physically, the topological zero temperature contribution to the fermion number corresponds to the virtual dipole pairs of vacuum polarization, which do not live long enough to thermalize. The nontopological finite temperature corrections correspond to the real $q \bar{q}$ pairs in the thermal plasma which are sensitive to the details of the single particle spectrum. 

This paper is organized as follows. In section II, we define the finite $T$ induced fermion number and review why it is generically nontopological. In Section III we use the derivative expansion to compute higher order $T$ dependent corrections to the winding number of the chiral field. As a nontrivial check we show that it vanishes in the $T \to 0$ limit. We conclude in section IV and discuss the implications of the existence of nontopological $T$ dependent corrections for the hybrid chiral bag model for the nucleon.

\section{Finite Temperature Induced Fermion Number}

The induced fermion number is an expectation value of the second quantized fermion number operator, $N=\frac{1}{2}\int d{\bf x}\, [\Psi^\dagger,\Psi]$. For a given static classical background field configuration, the second quantized fermion field operator $\Psi$ can be expanded in a complete set of states of the single particle Dirac Hamiltonian $H$. At zero temperature, the fermion number is a vacuum expectation value $\langle N\rangle_0\equiv\langle 0\vert N\vert 0\rangle$, and by a standard Fock space calculation \cite{niemi} is related to the spectral asymmetry of the Dirac Hamiltonian
\begin{eqnarray}
\langle N\rangle_0&=&-\frac{1}{2}\,(\textrm{spectral asymmetry})\nonumber\\
&=&-\frac{1}{2}\int_{-\infty}^{\infty} dE \, \sigma(E)
\,\,\textrm{sign}(E)
\label{asymmetry}
\end{eqnarray}
Here $ \sigma (E)$ is the spectral function, or density of states, of the Dirac Hamiltonian $H$ :
\begin{equation} 
\sigma (E)=\frac{1}{\pi}\textrm{ Im Tr}
\,\left(\frac{1}{H-E-i\epsilon}\right)
\label{spectral}
\end{equation}
Thus the fermion number of the vacuum $\langle N \rangle_0$ is essentially the difference between the number of positive and negative energy states of the Dirac spectrum. When fermions interact with topologically nontrivial backgrounds (solitons, vortices, monopoles, skyrmions), or are constrained by certain boundary conditions (e.g., in the chiral bag), the spectral asymmetry can be nonintegral, because of distortions in the Dirac sea \cite{jackiw,wilczek,niemi}.

 At nonzero temperature, $T$, the induced fermion number is a
{\it thermal} expectation value :
\begin{eqnarray}
\langle N\rangle_T &=&\frac{\textrm{Tr}\,(e^{-\beta
H}N)}{\textrm{Tr}\,(e^{-\beta H})}\nonumber\\
                   &=&-\frac{1}{2}\int_{-\infty}^{\infty}dE \,\sigma (E)
                   \,\tanh\left(\frac{\beta  E}{2}\right)
\label{nt}
\end{eqnarray}
where $\beta \equiv \frac{1}{T}$. This finite temperature
expression (\ref{nt}) reduces smoothly to the zero temperature expression
(1) in the zero temperature limit $\beta\to \infty$. The expression (\ref{nt}) also emphasizes the fact that the spectrum of single particle states for fermions interacting with the classical background field is completely determined by the single particle Dirac Hamiltonian $H$ of the system and has nothing to do with temperature. The temperature only influences the probability of each energy level being occupied. The finite temperature induced fermion number $\langle N\rangle _T$ splits naturally into a zero temperature piece $\langle N\rangle _0$, given by (\ref{asymmetry}) and a temperature dependent correction
\begin{equation}
\langle N\rangle _T=\langle N\rangle _0 +\int_{-\infty}^{\infty}dE \,\sigma (E) 
\,\textrm{sign}(E)\, n\,(|E|)\quad 
\label{split}
\end{equation}
 Here $n(E)$ is the Fermi-Dirac distribution function
\begin{equation}
n(E)=\frac{1}{e^{\beta E} +1}
\end{equation}
and we have used the simple identity  $\tanh (\beta E/2) =1 - 2 n(E)$. Thus the temperature enters only in the second term of (\ref{split}) via the Fermi-Dirac distribution $n(E)$. While $\langle N\rangle_0$ is topological, the finite temperature correction to the induced fermion number, being dependent on the Fermi weighting factor $n(|E|)$, is more sensitive to the details of the fermion spectrum, which in turn depends on the precise shape of the background field.  Therefore, the generic situation is that the finite temperature correction is 
nontopological. However, in certain special cases, when the background is such that the Dirac Hamiltonian has a quantum mechanical SUSY, then the odd part of the spectral function $\sigma (E)$ [which is what is needed in (\ref{split})] is itself topological, and therefore the finite temperature correction is also topological even though it is temperature dependent. The simplest cases where this happens are for a kink background in 1+1 dimensions \cite{semenoff,dunneian}, or for a monopole background \cite{cp,feinberg}.

The physical interpretation of the separation in (\ref{split}) is the following \cite{dunneian}. The zero temperature fermion number  $\langle N\rangle _0$ represents the vacuum polarization of the fermions in the presence of the background; it is temperature independent because the virtual dipoles do not live long enough to come to thermal equilibrium. The temperature dependent correction corresponds to the plasma response of the system in the presence of the background and is thus sensitive to the details of the spectrum. The finite temperature induced fermion number (and the rms fluctuations of the fermion number) have been computed for sigma model backgrounds in (1+1) and (2+1) dimensions \cite{dunneian,dunne,fluc,nonlinearsigma}. In this paper we consider second quantized fermions chirally coupled to an SU(2)  background in (3+1) dimensions.

\section{Induced Fermion Number in a Chiral Field}

\subsection{The Model}

Consider an isodoublet of fermions interacting in an $\textrm{SU}(2)_L \times \textrm{SU}(2)_R$ chiral invariant way with a classical, static chiral background in (3+1) dimensions
\begin{eqnarray}
\mathcal{L} &=& i \bar{\Psi} \gamma ^{\mu} \partial _{\mu} \Psi - \bar{\Psi} \left(\sigma + i \gamma _5 \vec{\tau} \cdot \vec{\pi} \right) \Psi \nonumber \\
&=& i \bar{\Psi} \gamma ^{\mu} \partial _{\mu} \Psi -m \bar{\Psi}\, \rm{exp} \left(i \gamma_5 \vec{\tau} \cdot \vec{\phi}\right)\Psi
\label{lagrangian}
\end{eqnarray}
Here $\sigma$ is a scalar field and $\vec{\pi}$ is a triplet of pseudoscalar (pion) fields. The Pauli matrices, $\vec{\tau}$, are the generators of SU(2), and $m$ is the mass scale in the theory. The Goldstone boson fields ($\sigma,\vec{\pi}$)  are  constrainted to lie on the ``chiral circle'',  $\sigma^2 +\vec{\pi} \cdot \vec{\pi} =m^2$, and thus they can be parametrized by a unitary matrix  $U$ as
\begin{equation}
U=\textrm{exp}\left(i \vec{\tau} \cdot \vec{\phi}\right)=\frac{1}{m}(\sigma +i \vec{\tau} \cdot \vec{\pi})
\label{unitary}
\end{equation}
Thus
\begin{eqnarray}
\sigma&=&m \cos (|\vec{\phi}|) \nonumber  \\
\vec{\pi}&=& m \sin (|\vec{\phi}|)\hat{\pi}
\label{phifield}
\end{eqnarray}
where the hat denotes a unit vector in isospin space.  The model (\ref{lagrangian}) is used in various quark-soliton models of baryons wherein the baryon is described in terms of quarks interacting with a soliton of the classical chiral field \cite{diakonov,ripka}. The model (\ref{lagrangian}) also finds applications in extensions of the MIT bag model of the nucleon, to incorporate chiral symmetry; in these ``hybrid'' chiral bag models \cite{jaffe,bhaduri},  quarks confined inside a boundary are chirally coupled to scalar and pseudoscalar fields (mesons) at the bag boundary with appropriate boundary conditions to maintain continuity of the axial vector current.

At zero temperature, the induced fermion number for the model (\ref{lagrangian}) can be computed using the derivative expansion technique and is given by \cite{gold,niemi,diakonov,dhoker}
\begin{equation}
\langle N\rangle _0 =\frac{1}{24 \pi ^2}\, \int d^3x\, \epsilon ^{ijk}\, \textrm{Tr} \left(\partial _{i}U U^{\dagger} \partial _{j}U U^{\dagger}  \partial_{k}U U^{\dagger}\right)
\label{topcharge}
\end{equation}
Thus, the zero temperature induced fermion number is simply the winding number of the group element $U({\bf x})$. The winding number is a topological invariant, determined by the asymptotic behavior of $U({\bf x})$.  Invoking Lorentz invariance, the topological charge density (the integrand of $\langle N\rangle _0$) can be interpreted as the zeroth component of a current density
\begin{equation}
\textrm{B} _{\mu} =\frac{1}{24 \pi ^2}\,  \epsilon _{\mu \nu \rho \sigma} \textrm{Tr} \left(\partial^{\nu}U U^{\dagger} \partial^{\rho}U U^{\dagger}  \partial^{\sigma}U U^{\dagger}\right)
\label{topcurrent}
\end{equation} 
This current $\textrm{B}_{\mu}$ is often called a topological current because it is conserved, $\partial_{\mu} \textrm{B}^{\mu}=0$,  without use of the equations of motion. 
In fact, the current density $\textrm{B}_{\mu}$ in (\ref{topcurrent}) has corrections involving higher order derivatives of the field $U$. However, these
higher derivative corrections are total derivatives and
thus do not contribute to an integrated quantity such as the fermion
number $\langle N\rangle _0$ \cite{wilczek,diakonov}. At nonzero temperature, 
experience with similar models in lower dimensions \cite{dunne,nonlinearsigma} suggests that the fermion number density $\textrm{B}_{0}$ receives higher order derivative corrections that are {\it not} total derivatives and that are nontopological. This expectation will be confirmed by an explicit calculation in the next section for the $3+1$ dimensional chiral quark model (\ref{lagrangian}). This appearance of nontopological corrections at finite temperature is completely consistent with the fact that the Dirac Hamiltonian for this system has no quantum mechanical SUSY. Indeed, numerical results \cite{ripka,kahana} for the spectrum of fermions in the background of an SU(2) chiral field $U$, in the ``hedgehog'' ansatz, show that the quark energy levels are highly sensitive to the shape (in particular, the length scale) of the hedgehog profile function. By the arguments of the previous section, this means that the finite temperature corrections in (\ref{split}) are necessarily  nontopological.
 
\subsection{The Derivative Expansion Calculation}

To calculate the induced fermion number at nonzero temperature we adopt the general methodology used in \cite{dunneian}. Using the $i \epsilon$ prescription in the spectral density (\ref{spectral}) we rewrite $\langle N\rangle _T$ as contour integral using the Dirac resolvent
\begin{equation}
R(z) \equiv \textrm{Tr}\left( \frac{1}{H-z}\right)
\label{resolvent}
\end{equation}
Here $z$ is an arbitrary complex number, and $H$ is the single particle Dirac Hamiltonian corresponding to (\ref{lagrangian}) in the fermion sector. Thus
\begin{equation}
\langle N\rangle _T=-\frac{1}{2}\int _{\mathcal{C}}\, \frac{dz}{2 \pi i} \, \textrm{Tr} \left(\frac{1}{H-z}\right) \tanh \left(\frac{\beta z}{2}\right)
\label{contour}
\end{equation}
where $\mathcal{C}$ is the contour $(-\infty +i \epsilon, +\infty +i
\epsilon)$ and $(+\infty -i \epsilon, -\infty -i \epsilon)$ in the
complex energy plane.  For a Hermitian Hamiltonian $H$, the poles and
branch cuts of the resolvent lie on the real axis. On the other hand,
the Matsubara poles $z_n=(2n+1)i \pi T$ of the $\tanh(\beta z/2)$ function lie on the imaginary axis. Thus we have two
equivalent representations for $\langle N\rangle _T$ :  deforming
the contour $\mathcal{C}$ around the poles and cuts of $R(z)$ leads to
an integral representation, and deforming $\mathcal{C}$ around the
discrete $z_n$ leads to a summation representation. The integral
representation is just the familiar Sommerfeld-Watson transform of the
summation representation. Thus the computational strategy is:

{\bf (i)} Obtain an expression for the Dirac resolvent $R(z)$.  In most cases this requires an approximation since the Dirac equation in the presence of complicated backgrounds can rarely be solved exactly, especially in higher dimensions. In this paper we use the derivative expansion to obtain an approximate expression for $R(z)$.

{\bf (ii)} Insert the resulting expression for $R(z)$ into (\ref{contour}) to obtain an expression for $\langle N \rangle _T$ either as an integral, or as a sum over the Matsubara frequencies $z_n$. 

We stress that these two steps are independent. In particular, the derivative expansion calculation has nothing to do with the temperature -- it is simply a means to obtain approximate information about the spectral properties of the Hamiltonian in the presence of the static background. 

In the derivative expansion approximation \cite{wilczek,dhoker,derivative} one assumes that
derivatives of the background field (here the chiral field $U$) are much smaller than the mass scale $m$ in the theory,
\begin{equation}
|\vec{\nabla} U| \ll m
\label{condition}
\end{equation}
Since we are considering a static background configuration all these derivatives are spatial derivatives.  So the background field is assumed to be approximately constant on the length scale of the fermion Compton wavelength. First, note that to compute $\langle N\rangle _T$ we need the {\it odd} part of the spectral density $\sigma(E)$, or equivalently the {\it even} part (in $z$) of the resolvent  $R(z)$. The even part of the resolvent (\ref{resolvent}) can be written as
\begin{equation}
\left[\textrm{Tr} \left(\frac{1}{H-z}\right) \right]_{\rm{even}} =\textrm{Tr} \left(H \frac{1}{H^2-z^2}\right)
\label{reven}
\end{equation}
Our strategy is to develop a systematic derivative expansion for this even part of the resolvent.
From the Lagrangian (\ref{lagrangian}), the corresponding Dirac Hamiltonian for the fermions in the chiral background $U$ is
\begin{eqnarray}
H&=&-i I \otimes \gamma^0 \vec{\gamma} \cdot \vec{\nabla} +m \cos (|\vec{\phi}|) I \otimes \gamma^0 +i m \vec{\tau}\cdot \hat{\pi} \otimes \gamma^0 \gamma^5 \sin (|\vec{\phi}|) \nonumber  \\
&=& -i I \otimes \sigma^1 \otimes \sigma^i \, \, \partial_i +\frac{m}{2} \left[(U+U^\dagger) \otimes \sigma^3 \otimes I +i (U-U^\dagger) \otimes \sigma^2 \otimes I\right]
\label{hamiltonian}
\end{eqnarray}
where we have used the definitions (\ref{unitary}) and (\ref{phifield}) and $I$ is the $2 \times 2$ identity matrix. The first term is the kinetic energy of the fermions, the second is the chiral invariant fermion-meson interaction. We choose the gamma matrix  convention as in \cite{niemi}: $\gamma^0=\sigma^3 \otimes I,\,  \gamma^i=i \sigma^2 \otimes \sigma^i,\,  \gamma^5= \sigma^1 \otimes I$. The square of the Hamiltonian (\ref{hamiltonian}) is 
\begin{eqnarray}
H^2&=&I \otimes I \otimes I \,(-\vec{\nabla}^2 +m^2) +\frac{m}{2} \left[i \, (\partial_i U -\partial_i U^{\dagger}) \otimes \sigma^3 \otimes \sigma^i - (\partial_i U +\partial_i U^{\dagger}) \otimes \sigma^2 \otimes \sigma^i\right] \nonumber  \\
& \equiv & H_{0}^2 +V
\end{eqnarray}
where the interaction term $V$ is
\begin{equation}
V=\frac{m}{2} \left[i \, (\partial_i U -\partial_i U^{\dagger}) \otimes \sigma^3 \otimes \sigma^i - (\partial_i U +\partial_i U^{\dagger}) \otimes \sigma^2 \otimes \sigma^i\right]
\end{equation}
and $H_0 ^{2}=(-\vec{\nabla}^2 +m^2)$ is the square of the free Dirac Hamiltonian. Since $V$ involves one factor of $m$ and one derivative of $U$, it is small compared to the natural scale $m^2$ of $H_{0}^2$, owing to the derivative expansion condition (\ref{condition}). Hence the even part of the resolvent (\ref{reven}) can be systematically expanded in powers of $V$:
\begin{equation}
\left[\textrm{Tr}\left(\frac{1}{H-z}\right)\right]_{\rm{even}}= -\text{Tr}( H\triangle V \triangle V \triangle V \triangle) + \textrm{Tr}(H\triangle V \triangle V \triangle V \triangle V \triangle)-\textrm{Tr}(H\triangle V \triangle V \triangle V \triangle V \triangle V \triangle) + \cdots
\label{trace}
\end{equation}
where the free propagator $\triangle$ is 
\begin{equation}
\triangle \equiv \frac{1}{-\vec{\nabla}^2 +m^2 -z^2}
\end{equation}
The trace in (\ref{trace}) involves a matrix trace over the Dirac indices, a trace over the internal isospin indices, and a functional trace. The leading term in this expansion involves three factors of $V$, since the Dirac and Isospin traces involving $H$ and either one or two factors of $V$ give zero. Note that in computing these traces the operators $H$ and $\triangle$ act on everything to the right of them. We thus need to know the action of $\triangle$ on the perturbation $V({\bf x})$. We make use use of the following operator identity:
\begin{eqnarray}
\frac{1}{-\vec{\nabla}^{2}+a^{2}} V({\bf x})&=&V({\bf x})\frac{1}{-\vec{\nabla}^{2}+
a^{2}} +2\left(\vec{\nabla}V({\bf x})\right)\cdot
\vec{\nabla}\frac{1}{(-\vec{\nabla}^{2}+a^{2})^{2}}+\left(\vec{\nabla}^{2}
V({\bf x})\right)\frac{1}{(-\vec{\nabla}^{2}+a^{2})^{2}}                                                                                                    \nonumber\\ & & + 4\left(\partial_{i} \partial_{j}  V({\bf x})\right)\partial_{i} \partial_{j}
\frac{1}{(-\vec{\nabla}^{2}+a^{2})^{3}}+4\left[\vec{\nabla}\left(\vec{\nabla}^{2} V({\bf x})\right)\right] \cdot
\vec{\nabla} \frac{1}{(-\vec{\nabla}^{2}+a^{2})^{3}}+\cdots
\label{spatial}
\end{eqnarray}
where the parentheses around the derivatives on the RHS of (\ref{spatial}) indicate that the gradient operator acts on $V({\bf x})$ only.  At any given order of the derivative expansion we collect together the required number of derivatives of the field $U$ (all possible combination of derivatives for a given order), as given by repeated application of the above identity, noting that $V({\bf x})$ itself already involves one derivative of $U$. 

The leading term in the derivative expansion involves three factors of $V$ from the first term in (\ref{trace}). This corresponds to all factors of $V$ pulled out, with $\triangle$ having no action on them (corresponding to the first term in (\ref{spatial})):
\begin{eqnarray}
\left[\textrm{Tr}\left(\frac{1}{H-z}\right)\right]^{(3)} _{\rm{even}}&=& -\textrm{Tr} (H \triangle V \triangle V \triangle V \triangle)^{(3)}\nonumber \\
&=&\frac{-m^4}{16 \pi (m^{2} -z^{2})^{\frac{5}{2}}} \int d^3 x \, \epsilon^{ijk} \, \textrm{Tr} \left(U \partial _i U^{\dagger} \partial _j U \partial _k U^{\dagger} \right)
\end{eqnarray}
Substituting this for $\langle N \rangle _T$ in (\ref{contour}), we perform the contour integral as a sum over Matsubara modes, and find the finite $T$ contribution at this order of the derivative expansion to be
\begin{eqnarray}
\langle N \rangle _T ^{(3)} &=&\left(\sum_{n=-\infty}^\infty \frac{m^4 T}{16 \pi \left[m^2 +((2n +1)\pi T)^2\right]^{\frac{5}{2}}} \right) \int d^3 x \, \epsilon^{ijk}\,  \textrm{Tr} \left(U \partial _i U^{\dagger} \partial _j U \partial _k U^{\dagger} \right) \nonumber \\
 & = & \left(\sum_{n=-\infty}^\infty \frac{3 \pi m^4 T}{2 \left[m^2 +((2n+1)\pi T)^2\right]^{\frac{5}{2}}} \right) \langle N\rangle _0
\label{nt3}
\end{eqnarray}
Notice that this leading-order term is simply the zero temperature fermion number $\langle N\rangle_0$, multiplied by a  temperature dependent prefactor. In the low temperature limit, this leading-order term (\ref{nt3}) reduces smoothly to the zero temperature fermion number $\langle N\rangle_0$ in (\ref{topcharge}):
\begin{equation}
\langle N\rangle _T ^{(3)}\sim \langle N\rangle _0 -\sqrt{\frac{\pi m^3}{2T^3}}e^{-\frac{m}{T}}\, \, \langle N\rangle _0 + \cdots
\label{leading}
\end{equation}
Here we have used the general low temperature expansion
\begin{equation}
T\sum_{n=-\infty}^\infty \frac{1}{(((2n+1)\pi T)^{2}+m^{2})^{p}} \sim 
\frac{m^{1-2p}}{2\sqrt{\pi}} \frac{\Gamma (p-\frac{1}{2})}{\Gamma
(p)}-\frac{(2mT)^{1-p}}{m\Gamma (p)} e^{-\frac{m}{T}}+ \cdots 
\label{identity}
\end{equation}
Thus, in the zero temperature limit the leading derivative expansion term involving three derivative of $U$ gives the entire zero temperature answer ({\ref{topcharge}), with temperature dependent corrections that vanish exponentially as $T\to 0$. 

The leading-order expression (\ref{nt3}) is still topological since it is simply the zero temperature answer multiplied by a smooth function of temperature.
The nontopological contributions to $\langle N\rangle _0$ come from the next higher term in the derivative expansion which involves five derivatives of $U$ (the fourth order terms in the derivative expansion give zero because the functional trace over the propagators turn out to be zero). At fifth order, each of the first three terms in the derivative expansion ({\ref{trace}) contributes.
Using the operator identity (\ref{spatial}) we collect together terms involving five derivatives of $U$ from each of the three terms in (\ref{trace}). For the last term involving five factors of $V({\bf x})$ this corresponds to taking all the $V$'s out with the $\triangle$'s having no effect on them. For the first two terms we need to go to higher derivative terms in (\ref{spatial}) to collect the five derivative terms. After a considerable amount of manipulation we have found the following compact expression for this fifth order contribution:
\begin{eqnarray}
\left[ \textrm{Tr}\left(\frac{1}{H-z} \right) \right]^{(5)} _{\rm{even}}
 &=& \frac{m^6 +6 m^4 z^2}{256 \pi (m^2 -z^2)^{\frac{9}{2}}}
\int d^3x \, \epsilon^{ijk} \, \textrm{Tr}\, \left[ \left\{ \partial _i U \partial _j U^{\dagger}(\vec{\nabla}U \cdot \vec{\nabla}\partial _k U^{\dagger})- \partial _i U^{\dagger} \partial _j U (\vec{\nabla}U^{\dagger} \cdot
  \vec{\nabla}\partial _k U)\right \} \right.
  \nonumber \\
& & \left. - 4\,\left \{ U \partial_i U^{\dagger}\partial_j U \partial_k
  U^{\dagger}(\vec{\nabla}U \cdot \vec{\nabla} U^{\dagger})\right \}
  \right] \nonumber \\
& & +\frac{ m^4}{256 \pi (m^2 -z^2)^{\frac{7}{2}}} \int d^3x \, \epsilon^{ijk} \, \textrm{Tr}\, \left[5 \vec{\nabla} \cdot \left \{\partial_iU^{\dagger} \partial_j U \partial_k U^{\dagger} \vec{\nabla}U-\partial_i U \partial_j U^{\dagger} \partial_k U \vec{\nabla}U^{\dagger}\right\} \right. \nonumber \\
& & -4 \partial_i \left\{ U \partial_j U^{\dagger} (\vec{\nabla}\partial_k U \cdot \vec{\nabla}U^{\dagger})-U \partial_j U^{\dagger}(\vec{\nabla}U \cdot \vec{\nabla}\partial_k U^{\dagger})\right \} \nonumber \\
& & -6 \partial_i \left \{ U^{\dagger} \partial_j U (\vec{\nabla}U^{\dagger} \cdot \vec{\nabla}\partial_k U)-U^{\dagger} \partial_j U (\vec{\nabla}\partial_k U^{\dagger} \cdot \vec{\nabla}U) \right \} \nonumber \\
& & -10 \left. \partial_i \left \{U \partial_j U^{\dagger} \partial_k U \vec{\nabla}^2 U^{\dagger}- U^{\dagger} \partial_j U \partial_k U^{\dagger} \vec{\nabla}^2 U\right \} \right]
\label{tracefive}
\end{eqnarray}
In obtaining (\ref{tracefive}) we have used the following identity
\begin{eqnarray}
 \textrm{Tr}\left[\vec{\nabla}^2 U \partial_iU^{\dagger} \partial_j U \partial_k U^{\dagger}-\vec{\nabla}^2 U^{\dagger} \partial_i U \partial_j U^{\dagger} \partial_k U\right] =
 -\textrm{Tr}\left[U \partial_i U^{\dagger} \partial_j U \partial_k U^{\dagger}(\vec{\nabla}U^{\dagger} \cdot \vec{\nabla}U)-U^{\dagger} \partial_i U \partial_j U^{\dagger} \partial_k U (\vec{\nabla}U \cdot \vec{\nabla}U^{\dagger})\right]
\nonumber
\end{eqnarray}
which can be easily proved using the unitarity of $U$, the cyclicity of the trace and the fact that $\vec{\nabla}U^{\dagger} \cdot \vec{\nabla}U=(\vec{\nabla} \sigma \cdot \vec{\nabla} \sigma + \vec{\nabla} \pi ^{a} \cdot \vec{\nabla} \pi ^{a}) I$, where $U$ is parametrized as in (\ref{unitary}). 

From (\ref{tracefive}) we see that the fifth-order term in the even part of the resolvent contains two types of contributions: first,
those that cannot be expressed as a total derivative (the first two terms), and second those which
are total derivatives. The total derivative terms go to zero after doing the spatial integrals. The
remaining terms in (\ref{tracefive}) are clearly nontopological and cannot be expressed
in terms of the algebraic structure of the winding number density in 
(\ref{topcharge}). These terms give a nonzero contribution
to the spatial integral and depend sensitively on the profile of the field
$U$, not just on the asymptotic behavior of $U$. Substituting the fifth-order resolvent expression (\ref{tracefive}) into the general expression (\ref{contour}) for the finite temperature fermion number, it is a simple matter to evaluate the $z$ integral. This  gives the fifth order contribution to the finite $T$ induced fermion number:
\begin{eqnarray}
 \langle N\rangle _T^{(5)}&=& - T \sum _{n=-\infty}^{\infty} \frac{m^6-6m^4((2n+1)\pi T)^2}{256 \pi \left[m^2 +((2n+1)\pi T)^2 \right]^{\frac{9}{2}}} \int d^3x \, \epsilon^{ijk}\, \textrm{Tr}\left[ \left\{ \partial _i U \partial _j U^{\dagger}(\vec{\nabla}U \cdot \vec{\nabla}\partial _k U^{\dagger})- \partial _i U^{\dagger} \partial _j U (\vec{\nabla}U^{\dagger} \cdot
  \vec{\nabla}\partial _k U)\right \} \right.
  \nonumber \\
& & \hskip 8cm\left. - 4\left \{ U \partial_i U^{\dagger}\partial_j U \partial_k
  U^{\dagger}(\vec{\nabla}U \cdot \vec{\nabla} U^{\dagger})\right \}
  \right] 
\label{result} 
\end{eqnarray}
Putting everything together the finite temperature induced fermion
number $\langle N\rangle _T$ of fermions coupled to a chiral field, up to fifth order in the derivative expansion, is the sum of (\ref{nt3}) and (\ref{result}):
\begin{eqnarray}
 \langle N\rangle _T &=& \left(\sum_{n=-\infty}^\infty \frac{3 \pi m^4
T}{2 \left[m^2 +((2n+1)\pi T)^2\right]^{\frac{5}{2}}} \right)
\frac{1}{24 \pi ^2}
 \int d^3 x \,\epsilon^{ijk}\, \textrm{Tr}\left(U
\partial _i U^{\dagger} \partial _j U \partial _k U^{\dagger} \right)
\nonumber \\
 & &\hskip -1cm - \left( T \sum _{n=-\infty}^{\infty} \frac{m^6-6m^4((2n+1)\pi
T)^2}{256 \pi \left[m^2 +((2n+1)\pi T)^2 \right]^{\frac{9}{2}}}\right)
 \int d^3x \, \epsilon^{ijk}\, \textrm{Tr}\left[ \left\{ \partial _i U
\partial _j U^{\dagger}(\vec{\nabla}U \cdot \vec{\nabla}\partial _k
U^{\dagger})- \partial _i U^{\dagger} \partial _j U
(\vec{\nabla}U^{\dagger} \cdot \vec{\nabla}\partial _k U)\right\}
\right. \nonumber \\
 & & \hskip 7cm- \left.  4\left \{ U \partial_i U^{\dagger}\partial_j U \partial_k
 U^{\dagger}(\vec{\nabla}U \cdot \vec{\nabla} U^{\dagger})\right \} \right] + \cdots 
\label{mainresult}
\end{eqnarray}
This is the main result of this paper. The first term is proportional to the (topological) winding number (\ref{topcharge}), and as shown in (\ref{nt3},\ref{leading}), the temperature dependent prefactor smoothly reduces to one (plus exponentially decaying corrections) as  $T\to 0$, and we recover the well-known zero temperature answer that the fermion number is equal to the winding number of the background field $U$. The second set of terms in (\ref{mainresult}) (coming from $\langle N\rangle _T ^{(5)}$), are clearly nontopological. A nontrivial consistency check is that this nontopological contribution must vanish in the $T\to 0$ limit, since the $T=0$ fermion number is topological and comes entirely from the third-order derivative expansion term.
This is shown as follows: using the low temperature expansion (\ref{identity}), the $T$ dependent prefactor of the nontopological piece in (\ref{mainresult}) behaves as (keeping leading and next-to-leading terms)
\begin{eqnarray}
- T\sum _{n=-\infty}^{\infty} \frac{m^6 -6m^4 ((2n+1)\pi T)^2}{256 \pi [m^2 +((2n+1)\pi T)^2]^{\frac{9}{2}}} & \sim & -\frac{m^4}{256 \pi} \left[ 7m^2\left(\frac{m^{-8}}{2\sqrt{\pi}}\, \frac{\Gamma(4)}{\Gamma(9/2)}-\frac{(2 mT)^{-7/2}}{m \Gamma(9/2)}\, e^{-m/T}+\dots\right)\right.
\nonumber\\
&&\hskip 1cm\left.
-6\left(\frac{m^{-6}}{2\sqrt{\pi}}\, \frac{\Gamma(3)}{\Gamma(7/2)}-\frac{(2 mT)^{-5/2}}{m \Gamma(7/2)}\, e^{-m/T}+\dots\right) \right]\nonumber \\
& \sim & \sqrt{\frac{2mT}{\pi}} \left(\frac{m}{3840 \pi T^4}\right)\left( 1-6 \frac{T}{m}+\cdots \right) e^{-\frac{m}{T}}
\label{cancel}
\end{eqnarray}
The important thing to note here is that the leading constant pieces cancel, leaving the leading contribution as an exponentially suppressed term which vanishes as $T\to 0$. This should be contrasted with the low temperature limit of the prefactor of the third order term in (\ref{nt3}), where a constant term survives, yielding the winding number, as in (\ref{leading}). The cancellation in (\ref{cancel}) of the leading constant pieces in the prefactor of the fifth order term in (\ref{mainresult}) means that this fifth order contribution (which is nontopological)  to the fermion number vanishes at $T=0$. This is a stringent check on the fact that the zero temperature fermion number is known to be topological. We stress that this cancellation is highly nontrivial because the compact fifth order expression (\ref{result}) comes about by combining  three different terms in the fifth order of the derivative expansion (\ref{trace}).  

\section{Conclusions and Discussion}
We have presented an explicit derivative expansion computation of the finite temperature corrections to the induced fermion number for fermions coupled to a static SU(2) chiral background. This calculation illustrates the splitting of the induced
fermion number into a zero temperature piece, which is topological (here it is the winding number of the chiral field), and a
finite temperature correction which is nontopological. That the finite
$T$ induced fermion  number will have a nontopological contribution
was argued on general grounds in \cite{fluc,nonlinearsigma}
where sigma models in (1+1) and (2+1) dimensions were analyzed. It was
shown in that generically there exist nontopological contributions to the induced
fermion number, except for very special background fields for which the Dirac Hamiltonian
has a quantum mechanical SUSY relating the positive and negative energy spectra.  The Dirac spectrum in the presence of a ``hedgehog''
chiral background has been calculated numerically in \cite{ripka,kahana} and clearly shows the absence of any quantum mechanical SUSY, as well as the sensitive dependence of the
spectrum on the scale of the background. Since the fermion spectrum is
not symmetric and is sensitive to the scale of the background (i.e., the length scale of the hedgehog profile function), the finite $T$ corrections must also be sensitive to this scale because
of the presence of the Fermi weighting factor $n(E)$ in
(\ref{split}). Thus the finite temperature corrections must be nontopological in this model. 
We have given an analytic calculation that confirms this numerical expectation and moreover yields the precise compact form (\ref{mainresult}) of the next order in the derivative expansion for the fermion number at finite temperature. 
Our calculation of the finite $T$ induced fermion
number was done in the derivative expansion limit which assumes that
the spatial derivatives of the background field are much smaller
than the fermion mass scale in the theory. A nontrivial consistency check on our result is that the nontopological finite $T$ contribution, coming from the fifth order in the derivative expansion, has a prefactor that vanishes exponentially in the $T \to 0$ limit.  

An interesting implication of our result is for the $\textrm{SU}_L (2) \times \textrm{SU}_R (2)$ hybrid chiral bag model of the nucleon. This model consists of quarks confined in a three dimensional spherical bag, of radius R, with appropriate boundary conditions so that the quark current vanishes outside the bag, ensuring confinement \cite{jaffe,bhaduri}. Inside the bag the quarks are free and at the bag surface they obey the chiral boundary condition
\begin{equation}
\left. \left[i \vec{\gamma} \cdot \hat{n} + \rm{exp}\left(i \vec{\tau}\cdot \hat{n} \gamma^{5} \theta \right)\right] \Psi\right]_{\rm bag\,\, boundary}=0
\label{bc}
\end{equation}
where $\Psi$ is the quark field and $\hat{n}$ is the unit normal vector at the bag boundary. From the point of view of inside the bag, the parameter $\theta$ is simply a boundary condition parameter. The boundary condition (\ref{bc})  is consistent with chiral symmetry in the sense that it ensures that the axial current is continuous across the bag surface. Goldstone and Jaffe computed the zero temperature fermion  number with $N_c =3$ quarks {\it inside} the bag \cite{jaffe}
\begin{equation}
\textrm{B}_{\rm{in}}(\theta)=1- \frac{1}{\pi}\left[ \theta-\frac{1}{2} \sin 2 \theta\right]
\label{inbag}
\end{equation}
The first term is simply the contribution of the valence quarks and the fractional piece comes from the contribution of the polarized Dirac sea. This polarization arises because the quark spectrum differs from the free spectrum owing to the nontrivial boundary condition (\ref{bc}). The topological nature of this zero temperature fermion number is illustrated by the fact that (\ref{inbag}) does not depend on the bag radius $R$.

In the hybrid chiral quark bag model \cite{jaffe,bhaduri}, on the outside of the bag the quarks are coupled to meson (Skyrme) fields, which are usually taken to be spherically symmetric, with the pseudoscalar pion fields in (\ref{phifield}) being in  the ``hedgehog ansatz'',  $\vec{\phi}=\hat{r} \theta(r)$. Outside the bag the quarks are coupled to a chiral field exactly as in the Lagrangian (\ref{lagrangian}). So the induced fermion number comming from the region  {\it outside} the bag can be calculated by computing the space integral in (\ref{topcharge}), using the hedgehog ansatz for $U$, from R to infinity \cite{jaffe}. This gives 
\begin{equation}
\textrm{B}_{\rm{out}}=\frac{1}{\pi}\left[ \theta (\rm{R})-\frac{1}{2} \sin 2 \theta (\rm{R}) \right] 
\label{outbag} 
\end{equation}
The topological nature of this contribution to the zero temperature fermion number is illustrated by the fact that (\ref{outbag}) only depends on the boundary values of the background field, not on its detailed shape. Next, the inside and outside regions are joined by identifying the boundary condition parameter $\theta$ with the value of the hedgehog profile field at the bag boundary: $\theta=\theta(R)$. 
Then combining (\ref{inbag}) and  (\ref{outbag}) we observe that the fractional parts of (\ref{inbag}) and (\ref{outbag}) cancel,  leaving the total fermion number of this hybrid chiral bag equal to one, the baryon number \cite{jaffe}. One can thus model the nucleon as a chiral bag defect in a Skyrme background. Clearly, this cancellation depends crucially on the topological nature of the induced fermion number, both inside and outside the bag.

We now ask whether this cancellation persists at nonzero temperature. From our results in this paper it is clear that the contribution from outside the bag, where the quarks are coupled to a chiral field, will be nontopological at nonzero temperature. The fifth-order term in (\ref{mainresult}), when integrated over the external region from $R$ to spatial infinity, clearly yields a temperature dependent contribution that is highly sensitive to the details of the profile function $\theta(r)$ appearing in the hedgehog field $U$, and not just to the boundary value $\theta(R)$. On the inside of the bag, the finite temperature induced fermion number is also nontopological. This follows from the fact that the detailed quark spectrum, with the boundary condition (\ref{bc}), is sensitive to the bag radius $R$. This can be deduced from  the explicit solutions in a grand spin basis (see \cite{farhi} for a phase shift analysis of the related external problem with boundary condition (\ref{bc})). Now suppose we keep the  identification $\theta=\theta(R)$ in order to ensure that the zero temperature cancellation occurs. Then we see immediately that this cancellation cannot persist at finite temperature -- this is because we always have the freedom to modify the profile function $\theta(r)$ in the external region, without changing its value at $r=R$. This will have no effect on the contribution from inside the bag because we have not changed $\theta$ [which is equal to $\theta(R)$], and we have not changed $R$. On the other hand, we see from (\ref{mainresult}) that the contribution from outside the bag does change. Therefore, the fractional parts of the internal and external contributions to the finite temperature fermion number cannot cancel.

Thus, at finite temperature, the baryon number of the hybrid chiral bag is no longer unity for arbitrary configurations of the chiral field $U$. Physically, this says that the thermal expectation value (\ref{nt}) of the baryon number differs from its topological zero temperature value of 1 (which is a temperature independent vacuum polarization effect) due to the thermal occupation of excited quark states (which is a temperature dependent plasma response effect). The finite temperature corrections are nontopological because they are more sensitive to the details of the quark spectrum, which in turn are highly sensitive to the precise form of the chiral background.

Finally, we mention that it would be interesting to investigate these issues analytically in the simpler cases of the (1+1) and (2+1) dimensional hybrid bags, where exact analytic information is available for the quark spectrum inside the bag \cite{zahed,wipf}.

\begin{acknowledgments}
We thank the U. S. Department of Energy for support through grant
DE-FG02-92ER40716.
\end{acknowledgments}


\end{document}